\documentclass[letterpaper]{article} 
\usepackage[preprint]{aaai2027}  
\usepackage[hyphens]{url}  
\usepackage{graphicx} 
\usepackage{natbib}  
\usepackage{caption} 
\usepackage{amssymb}

\usepackage{booktabs}
\usepackage{threeparttable}
\usepackage{makecell}
\usepackage{tabularx}
\usepackage{amsmath}
\usepackage{multirow}
\usepackage[caption=false]{subfig}
\usepackage{colortbl}
\usepackage{bm}
\definecolor{myblue}{RGB}{235, 245, 255}

\title{Prompt-Segmentation Meets ROI Encryption: Fine-Grained Video Protection Based on Diffusion Isolation}
\author{Xiang Zhang, Haoyan Lu, Zhenshan Tan, Ziqiang Li, Fei Peng, Zhangjie Fu}
\affiliations{}

\begin{document}

\maketitle

\begin{abstract}
Region-of-interest (ROI) selective encryption in H.265/HEVC aims to protect sensitive content while preserving the utility of non-ROI regions. Existing coding-unit (CU)-level approaches still suffer from substantial encryption redundancy for two reasons. First, existing approaches map object-detection bounding boxes to CUs, producing regular rectangular regions, which poorly match the shape of irregular ROI boundaries, such as facial contours. Second, their coding-tree-unit (CTU)-based diffusion isolation expands the protected footprint in fixed $64\times64$ blocks, causing encryption-induced distortion to spread over a large number of non-ROI pixels. To address the aforementioned shortcomings, we propose a prompt segmentation-guided CU-level ROI encryption algorithm, which mainly consists of two core components. The first component is a prompt segmentation-guided fine-grained ROI encryption pipeline. It maps pixel-level ROI masks to CUs according to an ROI-overlap threshold, refines boundary CUs to \(8\times8\) granularity, and applies format-compliant encryption to multiple syntax elements in the mapped ROI CUs. The second is a dual diffusion isolation strategy to address the encryption diffusion problem. For spatial isolation, the encoder refines non-ROI boundary regions into \(8\times8\) CUs and uses PCM mode to bypass intra-frame prediction. For temporal isolation, it removes motion-vector candidates whose reference blocks overlap ROI CUs. Experimental results demonstrate that the proposed method generates encryption regions that more closely match irregular ROI boundaries, effectively confines distortion to the ROI.
\end{abstract}

\section{Introduction}
Videos shared, stored, and analyzed across social media, surveillance, telemedicine, and other platforms often contain identifiable information, including faces and license plates. Unauthorized access to these regions creates privacy risks during video transmission, storage, and analysis~\cite{PADILLALOPEZ20154177}. Privacy-preserving video coding must therefore protect sensitive regions while preserving the visual utility of non-sensitive content.

Naive video encryption protects the entire bitstream but also conceals non-sensitive content and incurs unnecessary encryption overhead~\cite{5955103}. Selective encryption addresses this limitation by encrypting only designated syntax elements while preserving format compliance with the video coding standard~\cite{6316136}. As one of the most widely adopted video coding standards, H.265/HEVC has inspired extensive research on selective encryption. Among these approaches, ROI selective encryption \cite{dufaux2008scrambling,dufaux2008h,carrillo2008compression,unterweger2016building,zhang2018lightweight,hosny2022privacy,cho2024practical,farajallah2015roi,tew2016region,taha2018end,yu2023coding,11323309,li2024ppl,zhang2025visual} has attracted considerable attention because it further confines protection to sensitive regions while keeping the remaining visual content accessible.
\begin{figure}[t]
    \centering
    \includegraphics[width=1\linewidth]{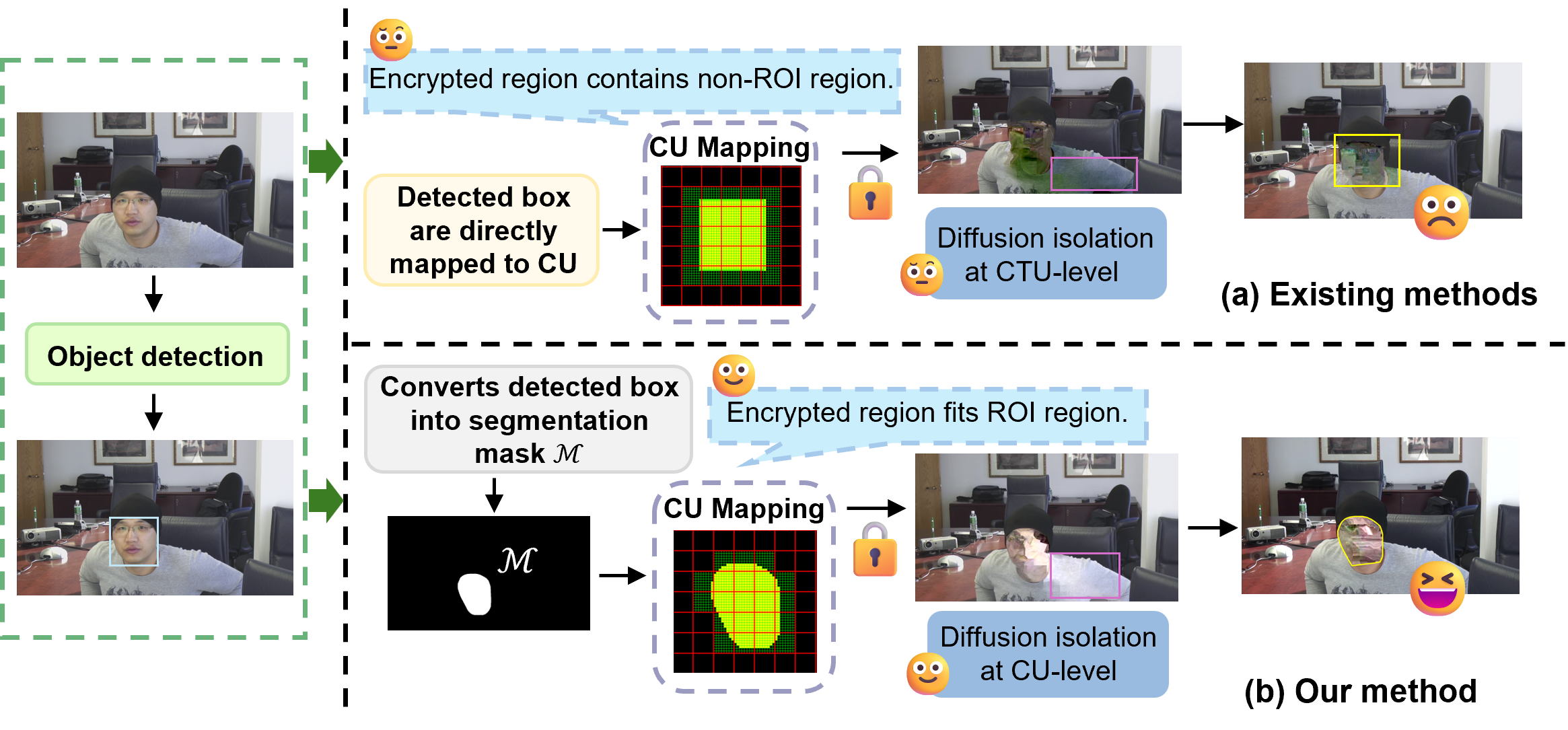}
    \caption{Comparison between existing methods and our method.}
    \label{fig:moti}
\end{figure}

Existing H.265/HEVC ROI encryption schemes operate primarily at Tile or CU level. Tile-based methods encrypt fixed rectangular regions, when a face occupies only part of a Tile, the surrounding background is inevitably encrypted together with the ROI~\cite{farajallah2015roi,tew2016region,taha2018end,zhang2025visual}. Although CU-based methods employ finer encryption units, they still suffer from two major limitations. First, as shown in Figure~\ref{fig:moti}(a), mapping rectangular detection boxes directly to CUs produces block-shaped regions that poorly fit irregular face contours, causing unnecessary background encryption~\cite{yu2023coding,11323309}. Second, existing diffusion isolation mechanisms rely on \(64\times64\)-sized CTUs, even with encryption units refined to the CU level, diffusion is only controlled within larger CTUs, causing encryption-induced distortion to spread to a large number of non-ROI.

To address these limitations, we propose a prompt segmentation-guided ROI selective encryption scheme with dual diffusion isolation. As shown in Figure~\ref{fig:moti}(b), the scheme first converts each detected ROI box into a pixel-level segmentation mask and labels CUs as ROI or non-ROI according to a mask-overlap threshold. Then, we encrypt selected intra prediction modes, motion information, and residual coefficients with an AES-CTR keystream while preserving valid H.265/HEVC syntax. Finally, we propose a dual diffusion isolation strategy based on PCM mode and MV restriction to address the encryption diffusion problem. The main contributions are as follows:

\begin{itemize}

\item \textbf{A prompt segmentation-guided fine-grained ROI encryption pipeline}: We propose a boundary-aware CU mapping mechanism combining with prompt segmentation to refine irregular ROI boundaries to \(8\times8\) granularity. Then, standard format-compliant selective encryption is applied to intra-prediction modes, motion information, and residual coefficients in the mapped ROI CUs. This design improves contour alignment and reduces unnecessary background encryption while preserving valid H.265/HEVC syntax.
\item \textbf{A dual diffusion isolation strategy based on PCM mode and MV restriction}: Our diffusion isolation strategy contains spatial and temporal isolation. For spatial isolation, we refine non-ROI CUs that may reference encrypted ROI CUs to an \(8\times8\) size and encodes them in PCM mode, preventing the spatial propagation of encryption-induced distortion. For temporal isolation, we discard motion-vector candidates whose reference blocks overlap ROI CUs, preventing temporal propagation.
\item \textbf{Extensive experiments have verified the effectiveness of the encryption algorithm}: We evaluate the proposed scheme on 35 sequences spanning resolutions from \(176\times144\) to \(4096\times2160\) against four representative Tile- and CU-level baselines. The scheme achieves the highest IoU and lowest ERR in all seven resolution groups and restores non-ROI PSNR from 32.86 to 42.92 dB through diffusion isolation mechanism.
\end{itemize}

\section{Related Works}

\subsection{Tile-level ROI Selective Encryption}
Tile-level ROI encryption represents the ROI as a set of rectangular H.265/HEVC Tiles. ~\cite{taha2018end} divide each frame into independent rectangular H.265/HEVC Tiles. At the CABAC level, a chaos-based keystream encrypts motion-vector differences and signs together with transform coefficients and their signs through XOR operations. The method also encrypts luma and chroma intra prediction modes through circular shifts within mode sets that preserve their coefficient-scanning directions. ~\cite{zhang2025visual} introduced a tunable framework. Their basic level encrypts MVD signs and values, residual signs and values, the Delta QP sign, Merge index, and the RefFrmIdx suffix. The enhanced level adds luma and chroma IPMs, MVPIdx, and the Delta QP value, while the advanced level applies Canny edge detection and chaos-based permutation to scramble nonzero coefficients in edge TUs. However, a Tile remains a fixed rectangular mapping unit. When a face occupies only part of a Tile or crosses Tile boundaries, the selected Tiles include surrounding non-ROI pixels. Therefore, Tile-level encryption cannot closely follow irregular ROI contours, this spatial mismatch leads to over-encryption, unnecessarily obscuring non-sensitive content and limiting the attainable trade-off between privacy protection and visual utility.

\subsection{CU-level ROI Selective Encryption}
CU-level methods map ROI locations to coding units and encrypt selected syntax elements, thereby providing finer-grained ROI protection than tile-level methods. ~\cite{yu2023coding} determine ROI CUs from the overlap between object-detection boxes and CU coordinates, encrypt multiple CABAC syntax elements, and constrain intra and inter prediction references. However, these reference constraints operate at a $64\times64$ CTU granularity, resulting in the final encrypted region still containing non-ROI region pixels that should not have been encrypted. ~\cite{11323309} map detection boxes to adaptively partitioned CUs and encrypt selected syntax elements using AES-CTR and XORShift32 to reduce encryption and encoding overhead. Their scheme lacks explicit intra and inter prediction isolation, leading to encryption spreading into non-ROI areas. In summary, despite their finer encryption granularity, existing CU-level methods still suffer from significant encryption redundancy, which remains a major challenge.

\section{Proposed Scheme}\label{sec3}
\begin{figure}[h]
    \centering
    \includegraphics[width=1\linewidth]{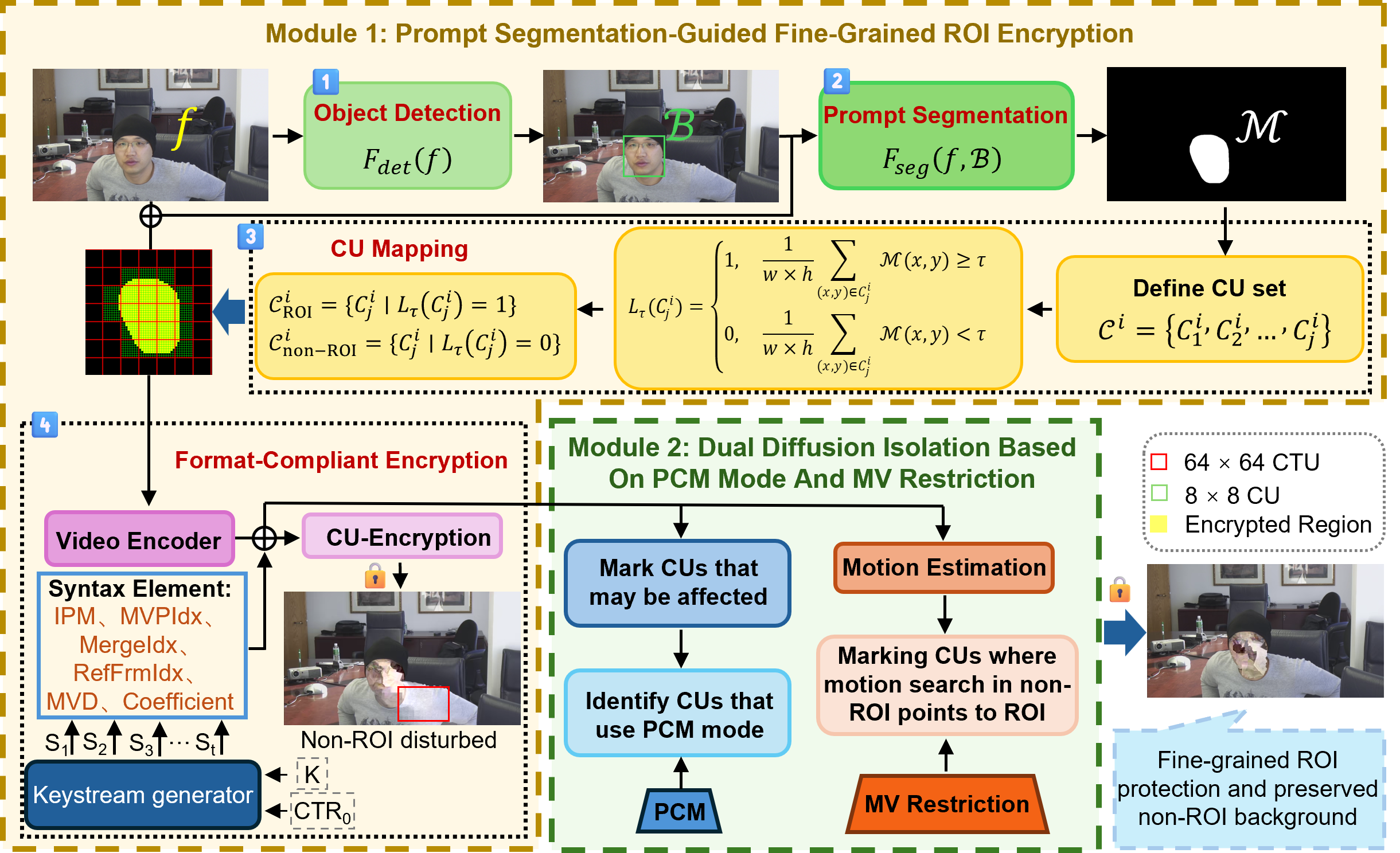}
    \caption{The proposed encryption scheme.}
    \label{fig:framework}
\end{figure}

Figure~\ref{fig:framework} shows the proposed scheme, which comprises two modules: prompt segmentation-guided fine-grained ROI encryption and dual diffusion isolation based on PCM mode and MV restriction. The first module detects sensitive objects, converts their bounding boxes into pixel-level masks, maps these masks to ROI CUs, and encrypts selected syntax elements while preserving H.265/HEVC format compliance. The second module confines distortion propagation through PCM mode and MV restriction, respectively. The specific implementation steps are described below.

\subsection{Prompt Segmentation-Guided Fine-Grained ROI Encryption}\label{sec3a}

Prompt segmentation-guided fine-grained ROI encryption comprises four components: object detection, prompt segmentation, CU mapping, and format-compliant encryption.
\subsubsection{Object Detection}
For the input video frame $f$, Through the detection function $F_{det}(\cdot)$, it predicts and locates the regions of interest within the video frame, and outputs the set of candidate object boxes for the current frame $\mathcal{B}$:
\begin{equation}
    \mathcal{B}=F_{det}(f).
\end{equation}

\subsubsection{Prompt Segmentation}

After obtaining the target bounding box $\mathcal{B}$ of the video frame, traditional rectangular boxes cannot accurately describe the edges of complex targets. Therefore, we use the target bounding box $\mathcal{B}$ as a spatial prior prompt to guide the segmentation model to perform precise pixel-level segmentation on the current video frame. Through the prompt segmentation function $F_{seg}(\cdot)$, a pixel-level mask corresponding to the target region of the current video frame is generated:
\begin{equation}
    \mathcal{M}=F_{seg}(f,\mathcal{B}).
\end{equation}

\subsubsection{CU Mapping}
After obtaining the pixel-level ROI mask, we propose a CU mapping rule to map the mask into the H.265/HEVC coding structure. Let the CU set of frame \(i\) be:
\begin{equation}
    \mathcal{C}^{i}=\{C^{i}_1,C^{i}_2,\ldots,C^{i}_j\},
\end{equation}
where \(C^{i}_j\) denotes the \(j\)-th CU in frame \(i\), with width \(w\), height \(h\). For each CU, we calculate the proportion of pixels covered by the predicted ROI mask \(\mathcal{M}\), $\mathcal{M}(x,y)$ represents the value of the ROI mask at pixel coordinates $(x,y)$. Given a mapping threshold \(\tau\in(0,1]\) (we will conduct a sensitivity analysis on the threshold $\tau$ and select the relatively optimal threshold in the experimental section), the CU-level ROI label is defined as:
\begin{equation}\label{threshold}
L_{\tau}(C^{i}_j)=
\begin{cases}
1, &
\displaystyle
\frac{1}{w\times{h}}
\sum_{(x,y)\in C^{i}_j}
\mathcal{M}(x,y)
\geq \tau,\\
0, &
\displaystyle
\frac{1}{w\times{h}}
\sum_{(x,y)\in C^{i}_j}
\mathcal{M}(x,y)
< \tau,
\end{cases}
\end{equation}
where \(L_{\tau}(C^{i}_j)=1\) indicates that the ROI coverage ratio of \(C^{i}_j\) reaches the threshold and that the CU is encrypted as an ROI-CU. Otherwise, \(L_{\tau}(C^{i}_j)=0\), and the CU is treated as a non-ROI CU. A smaller \(\tau\) favors ROI coverage, whereas a larger \(\tau\) reduces unnecessary encryption of non-ROI pixels. Accordingly, the ROI and non-ROI CU sets of frame \(i\) are:
\begin{equation}
\begin{aligned}
\mathcal{C}^{i}_{\mathrm{ROI}}=
 \{C^{i}_j\mid L_{\tau}(C^{i}_j)=1\},\\ 
\mathcal{C}^{i}_{\mathrm{non\text{-}ROI}}=
\{C^{i}_j\mid L_{\tau}(C^{i}_j)=0\}.
\end{aligned}
\end{equation}

\subsubsection{Format-Compliant Encryption}

After ROI mapping, we apply standard format-compliant selective encryption exclusively to the ROI CUs \(\mathcal{C}_{\mathrm{ROI}}^i\) in frame \(i\)~\cite{farajallah2015roi,tew2016region,taha2018end}. The encryption covers intra-prediction modes, motion information, and residual coefficients using an AES-CTR keystream, as shown in Table~\ref{tab:syntax_encryption}. For the $t$-th syntax element to be encrypted, AES-CTR generates a keystream fragment as:
\begin{equation}
S_t=\operatorname{AES}_{K}(CTR_0+t),
\end{equation}
where $K$ is the secret key and $CTR_0$ is the initial counter, $t$ is the sequence index of the encrypted syntax field. For finite-valued indices, The encoder uses a cyclic permutation to encrypt them as: 
\begin{equation}
P_{|\mathcal{R}|}(x,S_t)=\left(x+(S_t\bmod |\mathcal{R}|)\right)\bmod |\mathcal{R}|,
\end{equation}
where $x$ denotes the ordinal index of the original syntax value in its set of valid candidates $\mathcal{R}$, and $|\mathcal{R}|$ denotes the number of valid candidates. The value $S_t\bmod|\mathcal{R}|$ specifies a keystream-dependent cyclic shift. For binary or fixed-length fields, The XOR operation is used:
\begin{equation}
X_b(x,S_t)=x\oplus(S_t\bmod 2^b),
\end{equation}
where $b$ denotes the coded length of the syntax field in bits, and $S_t\bmod2^b$ provides a $b$-bit keystream mask.

\begin{table}[h!]
\small
\centering
\setlength{\tabcolsep}{5pt}
\renewcommand{\arraystretch}{1.05}
\begin{threeparttable}
\begin{tabular}{p{0.18\columnwidth}p{0.69\columnwidth}}
\toprule
\textbf{Operation} & \textbf{Syntax elements} \\
\midrule
$P_3(\cdot)$ &
Luma MPM index; chroma IPM index; RefFrmIdx with three active reference frames. \\
\midrule
$P_5(\cdot)$ &
MergeIdx. \\
\midrule
$X_1(\cdot)$ &
MVPIdx; RefFrmIdx with two active reference frames; horizontal and vertical MVD signs; residual coefficient signs; each selected coefficient-suffix bit. \\
\midrule
$X_2(\cdot)$ &
RefFrmIdx with four active reference frames. \\
\midrule
$X_5(\cdot)$ &
Non-MPM luma directional offset. \\
\bottomrule
\end{tabular}
\begin{tablenotes}[flushleft]
            \footnotesize
            \item[] \textit{Note:} $P_3(\cdot)$ and $P_5(\cdot)$ denote cyclic permutations over three and five valid indices, respectively. $X_1(\cdot)$, $X_2(\cdot)$, and $X_5(\cdot)$ denote XOR operations on one-, two-, and five-bit fields, respectively.
        \end{tablenotes}

\end{threeparttable}
\caption{Encryption operations for ROI CUs.}
\label{tab:syntax_encryption}
\end{table}

\subsection{Dual Diffusion Isolation Based on PCM Mode and MV Restriction}

Although selective encryption modifies only syntax elements associated with ROI CUs, the induced distortion may propagate to non-ROI regions through the prediction dependencies of H.265/HEVC. Specifically, intra prediction introduces spatial dependencies through reconstructed neighboring samples, while inter prediction establishes temporal dependencies through motion compensated reference blocks. For detailed analysis, please refer to the section 1 of \textbf{supplementary materials}. Therefore, diffusion isolation is formulated as a prediction dependency constraint problem. 

\subsubsection{Spatial Diffusion Isolation}

Intra prediction propagates ROI distortion only to non-ROI CUs whose prediction samples reference reconstructed samples from an ROI CU~\cite{2013Intra}. For an ROI CU $C^{i}_n$, let $\mathcal{D}_{\mathrm{intra}}(C^{i}_n)$ denote the set of candidate CUs located to its right, below, lower-right, upper-right, and lower-left. The vulnerable non-ROI CU set in frame $i$ is defined as
\begin{equation}
\begin{aligned}
\mathcal{V}^{i}
&=
\left\{
C^{i}_k\in\mathcal{C}^{i}_{\mathrm{non\text{-}ROI}}
\;\middle|\;
\substack{
\exists C^{i}_n\in\mathcal{C}^{i}_{\mathrm{ROI}},\\
C^{i}_k\in\mathcal{D}_{\mathrm{intra}}(C^{i}_n)
}
\right\}.
\end{aligned}
\end{equation}

We partition each CU in \(\mathcal{V}^{i}\) into \(8\times8\) CUs. Among these partitioned CUs, non-ROI \(8\times8\) CUs located within the five intra prediction reference neighborhood of ROI CUs are encoded in PCM mode, while the remaining non-ROI \(8\times8\) CUs retain the standard H.265/HEVC coding process:
\begin{equation}
\mathcal{C}^{i}_{\mathrm{PCM}}
=
\left\{\hat{C}\in\operatorname{Split}_{8\times8}\!\left(\mathcal{V}^{i}\right)\ \middle|\
\substack{\exists C^{i}_n\in\mathcal{C}^{i}_{\mathrm{ROI}},\ \hat{C}\sim C^{i}_n,\\
w=h=8,\ \operatorname{Mode}(\hat{C})=\mathrm{PCM}}\right\},
\end{equation}
where, $\operatorname{Split}_{8\times8}(\cdot)$ denotes CU partitioning to $8\times8$ granularity, \(\hat{C} \sim C_n^i\) indicates that the \(8 \times 8\) non-ROI CU \(\hat{C}\) is located within the five intra prediction reference neighborhood of the ROI CU \(C_n^i\), and \(\operatorname{Mode}(\cdot)\) denotes the coding mode assigned to a CU. PCM mode skips intra prediction mode and directly encodes the original sample values. Therefore, PCM-coded boundary CUs no longer reference distorted ROI reconstructions, blocking spatial distortion propagation from the ROI into non-ROI regions.

\subsubsection{Temporal Diffusion Isolation}

Inter prediction propagates encryption distortion when a non-ROI CU uses a motion candidate whose reference block intersects the ROI footprint in a reference frame. Let $\mathcal{F}^{i}$ is the set of reference frames available to frame $i$, and $r\in\mathcal{F}^{i}$ is one reference frame. The non-ROI CUs in frame $i$ exposed to distortion diffusion are:
\begin{equation}
\mathcal{C}_{\mathrm{risk}}^{i}
=
\left\{
C_m^i\in\mathcal{C}^{i}_{\mathrm{non\text{-}ROI}}
\;\middle|\;
\substack{
\exists r\in\mathcal{F}^{i},\ \exists\bm{k}\in\mathcal{K}^{i\rightarrow r}(C_m^i),\\
Bk^r(C_m^i,\bm{k})\cap\mathcal{A}^{r}\neq\varnothing
}
\right\},
\end{equation}
where \(\mathcal{K}^{i\rightarrow r}(C_m^i)\) denotes the set of candidate motion vectors for predicting the non-ROI CU \(C_m^i\) in the current frame \(i\) from reference frame \(r\). Each candidate $\bm{k}$ in \(\mathcal{K}^{i\rightarrow r}(C_m^i)\) specifies a motion vector that locates a reference block $Bk^r(C_m^i,\bm{k})$ in frame \(r\). $\mathcal{A}^{r}$ denotes the pixel area occupied by all ROI CUs in frame $r$. For each CU $C_p^i$ in $\mathcal{C}_{\mathrm{risk}}^{i}$, we discard candidate motion vectors whose reference blocks intersect the ROI footprint. The resulting safe candidate set without diffusion for the reference frame $r$ is:
\begin{equation}
\mathcal{K}^{i\rightarrow r}_{\mathrm{safe}}(C_p^i)
=
\left\{
\bm{k}\in\mathcal{K}^{i\rightarrow r}(C_p^i)
\;\middle|\;
Bk^r(C_p^i,\bm{k})\cap\mathcal{A}^{r}=\varnothing
\right\}.
\end{equation}

We perform inter-mode selection jointly over the safe candidate sets \(\mathcal{K}^{i\rightarrow r}_{\mathrm{safe}}(C_p^i)\) for all \(r\in\mathcal{F}^{i}\). This restriction prevents inter prediction from transferring ROI distortion into non-ROI.

\section{Experiments}

\subsection{Experimental Setup}
\paragraph{Implementation and Coding Configuration.}
We implement the proposed scheme in the H.265/HEVC reference software HM 16.9.
We run all experiments on a workstation equipped with an Intel Core i7-9750H CPU, 16 GB of memory, and Windows 11.
The software environment includes Microsoft Visual Studio 2010, MATLAB R2021a, and OpenCV 2.4.7.
We use the \texttt{encoder\_lowdelay\_main} configuration with an IBBBBB Group of Pictures (GOP) structure.
We enable PCM mode and set both \texttt{PCMLog2MinSize} and \texttt{PCMLog2MaxSize} to 3, restricting PCM coding to \(8\times8\) CUs.
We evaluate all methods at QP values of 8, 24, and 40.

\paragraph{ROI Detection and Segmentation.}
We treat faces as the sensitive ROIs in all experiments.
We train YOLOv8~\cite{yolov8} on WIDER FACE for face detection and use each predicted face box as a spatial prompt for SAM.
We use the ViT-H variant of SAM~\cite{kirillov2023segment} with the \texttt{vit\_h} checkpoint.
SAM converts each prompt into a pixel-level face mask, which the proposed mapping method subsequently projects onto the H.265/HEVC CU structure.

\paragraph{Test Sequences.}
We evaluate 35 video sequences covering resolutions from \(176\times144\) to \(4096\times2160\). We organize the sequences into seven resolution groups, with five sequences in each group. The test set includes single-person close-ups, multi-person scenes, indoor conversations, outdoor pedestrians, complex backgrounds, and fast-motion content. We set the frame rate to 60 fps and evaluate 100 frames from each sequence. Unless otherwise specified, all quantitative results are averaged across the three QP settings (8, 24, and 40) and the first 100 frames of every sequence in the seven resolution groups. Table~\ref{tab2} lists the complete test set.

\begin{table}[h]
\centering
\small
\setlength{\tabcolsep}{2pt}
\renewcommand{\arraystretch}{1.08}
\begin{tabularx}{\columnwidth}{
@{}
>{\centering\arraybackslash}p{0.20\columnwidth}
>{\raggedright\arraybackslash}X
>{\centering\arraybackslash}p{0.23\columnwidth}
@{}
}
\toprule
Group & Sequence & Resolution \\
\midrule
Group A &
Carphone / Claire / Foreman / Salesman / Suzie &
\(176\times144\) \\
Group B &
Akiyo / Crew / Mother\_Daughter / Paris / Silent &
\(352\times288\) \\
Group C &
BasketballDrill / BQMall / Keiba / PartyScene / RaceHorses &
\(832\times480\) \\
Group D &
FourPeople / Johnny / Vidyo1 / Vidyo3 / Vidyo4 &
\(1280\times720\) \\
Group E &
BasketballDrive / DinnerScene / Kimono / MeridianTalk / Tennis &
\(1920\times1080\) \\
Group F &
Beauty / FlowerKids / Jockey / ReadySteadyGo / RaceNight &
\(3840\times2160\) \\
Group G &
BoxingPractice / Crosswalk / Dancers / Narrator / Tango &
\(4096\times2160\) \\
\bottomrule
\end{tabularx}
\caption{Test video sequences used in the experiments.}
\label{tab2}
\end{table}

\paragraph{Baselines and Comparison Protocol.}
In order to comprehensively evaluate the performance of the proposed ROI selective encryption scheme in various aspects, We adopt the ROI-encryption evaluation benchmark introduced in~\cite{zhang2025visual} to assess localization accuracy and ROI distortion. For detailed evaluation methods, please refer to literature \cite{zhang2025visual}. We compare the proposed method against the tile-level schemes of ~\cite{taha2018end} and ~\cite{zhang2025visual}, and the CU-level schemes of ~\cite{yu2023coding} and ~\cite{11323309}. All methods process the same sequences and frame indices using identical QP values, GOP structures, rate-control settings, and decoding configurations. Each method receives the same YOLOv8 detection boxes and converts them into its native tile- or CU-level encryption units.

\subsection{Sensitivity analysis of overlap threshold \(\tau\)}

The \(\tau\) mentioned in Eq.~\ref{threshold} determines whether a CU belongs to the encrypted ROI according to its coverage by the input segmentation mask. A lower threshold includes more boundary CUs, improving ROI coverage while encrypting additional non-ROI pixels. A higher threshold produces a tighter encrypted region but may omit ROI pixels near object boundaries. To evaluate the sensitivity of the threshold \(\tau\), we select the first sequence from each video group as validation sequences for threshold selection and evaluate different values of \(\tau\) ranging from 0.1 to 1.0 with a step size of 0.02. IoU and bitrate change are used as evaluation metrics to assess the performance under different \(\tau\). Among them, IoU is a metric introduced by \cite{zhang2025visual}, which measures the spatial consistency between the final encrypted CU set generated by the encoder and the pixel-level ground-truth facial ROI mask obtained by manually delineating facial contours. 

It can be seen from the figure that IoU increases rapidly at low thresholds and then remains between approximately 0.87 and 0.90 over a broad intermediate range. This trend indicates that increasing \(\tau\) initially removes redundant non-ROI pixels without excluding many ROI pixels. The IoU reaches its maximum value of 0.8988 at \(\tau=0.84\). Beyond this point, stricter mapping excludes more boundary pixels and reduces IoU. The bitrate change generally decreases as \(\tau\) increases, although variations in the selected ROI-CU layout produce several local fluctuations. At \(\tau=0.84\), the bitrate increases only by 9.43\%. Therefore, \(\tau\)=0.84 is adopted in all subsequent experiments because it achieves the highest ROI-mapping IoU while maintaining a limited bitrate overhead.

\begin{figure}[h]
    \centering
    \includegraphics[width=\linewidth]{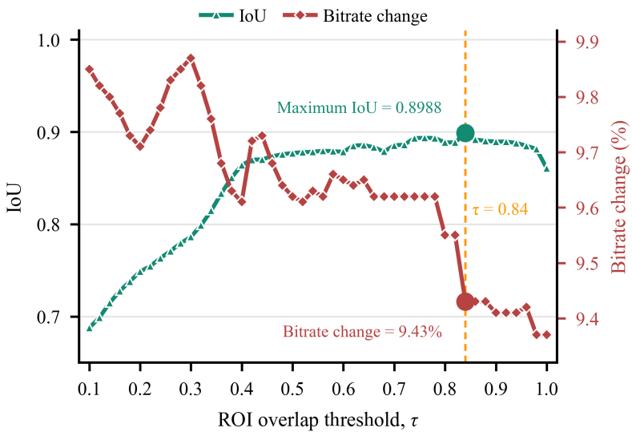}
    \caption{Impact of overlap threshold on IoU and bitrate}
    \label{th}
\end{figure}

\subsection{Comparison of ROI encryption accuracy}
\paragraph{Accuracy of Spatial Positioning.}
In order to quantitatively evaluate the accuracy of the proposed ROI encryption scheme in spatial positioning, we compared the IoU of five encryption methods. Table~\ref{tab:IoU} shows the average values of the comparison results for each group of video sequences. Our method achieved the highest IoU in all seven sequence groups, ranging from 0.767 to 0.957 (Table~\ref{tab:IoU}). Existing methods rely on tile partitions or box-to-CU mapping. The resulting regular regions include more background pixels, reducing their overlap with the ground-truth ROI. However, prompt segmentation and $8\times8$ size boundary CUs allow our method to follow ROI contours more closely, producing the highest IoU.
\begin{table}[h]
\centering
\small
\centering 
\setlength{\tabcolsep}{8pt}
\begin{tabular}{lccccc} 
\toprule
\multirow{2}{*}{Group} & \multicolumn{5}{c}{IoU} \\ 
\cmidrule{2-6} 
 & Taha & Zhang & Yu & Amael & Ours \\ 
\midrule
Group A & 0.142 & 0.283 &  0.723 & 0.723  &  \textbf{0.767} \\
Group B & 0.183 & 0.246 &  0.257 & 0.257  &  \textbf{0.859} \\
Group C & 0.167 & 0.416 &  0.433 & 0.443  &  \textbf{0.863} \\
Group D & 0.247 & 0.546 &  0.583 & 0.583  &  \textbf{0.891} \\
Group E & 0.228 & 0.721 &  0.702 & 0.702  &  \textbf{0.934}\\
Group F & 0.313 & 0.746 &  0.746 & 0.746  &  \textbf{0.953}\\ 
Group G & 0.239 & 0.801 &  0.778 & 0.778  &  \textbf{0.957} \\ 
\bottomrule
\end{tabular}
\caption{Average IoU of the compared schemes}
\label{tab:IoU}
\end{table}

\paragraph{Redundancy Encryption Ratio.}
To further evaluate the spatial granularity of different ROI encryption schemes, we additionally introduces Encryption Redundancy Rate (ERR) to measure the proportion of non-ROI that are unnecessarily included in the actual encrypted area, thereby reflecting how accurately the encrypted region fits the ground-truth ROI. Let $G$ denote the pixel-level ground-truth facial ROI mask obtained by manually delineating facial contours and $E$ denote the pixel footprint of the final encrypted CUs. The encryption redundancy rate is defined as:
\begin{equation}
    \text{ERR} = \frac{\lvert \text{E} \rvert - \lvert \text{E} \cap \text{G} \rvert}{\lvert \text{E} \rvert}
\end{equation}
where $\lvert \text{E} \rvert$ represents the area of the actual encrypted region, and $ \lvert \text{E} \cap \text{G} \rvert$ denotes the overlapping area between the actual encrypted region and the ground-truth ROI. A lower ERR indicates that the encrypted region is closer to the true ROI, with higher spatial precision. The comparison results are shown in Table~\ref{tab:ERR}. It can be seen from Table~\ref{tab:ERR} that our method achieved the lowest ERR across all seven groups (0.004--0.039). The final encryption result of existing methods is still constrained by large rectangular blocks, which increases redundancy. By contrast, pixel-level masks and $8\times8$ boundary refinement exclude most background pixels before encryption, producing the lowest ERR.

\begin{table}[h]
\centering
\small
\setlength{\tabcolsep}{8pt}
\begin{tabular}{lccccc} 
\toprule
\multirow{2}{*}{Group} & \multicolumn{5}{c}{ERR} \\ 
\cmidrule{2-6} 
 & Taha & Zhang & Yu & Amael & Ours \\ 
\midrule
Group A & 0.754 & 0.617 &  0.173 & 0.173  &  \textbf{0.039} \\
Group B & 0.711 & 0.622 &  0.634 & 0.634  &  \textbf{0.005} \\
Group C & 0.736 & 0.484 &  0.496 & 0.496  &  \textbf{0.027} \\
Group D & 0.741 & 0.354 &  0.326 & 0.326  &  \textbf{0.023} \\
Group E & 0.673 & 0.271 &  0.179 & 0.179  &  \textbf{0.006}\\
Group F & 0.583 & 0.173 &  0.138 & 0.138  &  \textbf{0.004}\\ 
Group G & 0.651 & 0.198 &  0.112 & 0.112  &  \textbf{0.004} \\ 
\bottomrule
\end{tabular}
\caption{Average ERR of the compared schemes}
\label{tab:ERR}
\end{table}

\subsection{Comparison of ROI perturbation effects}
\begin{figure*}[h]
\centering
\subfloat[Original]{
\includegraphics[width=0.150\textwidth]{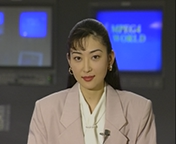}}
\hfill
\subfloat[Taha]{
\includegraphics[width=0.150\textwidth]{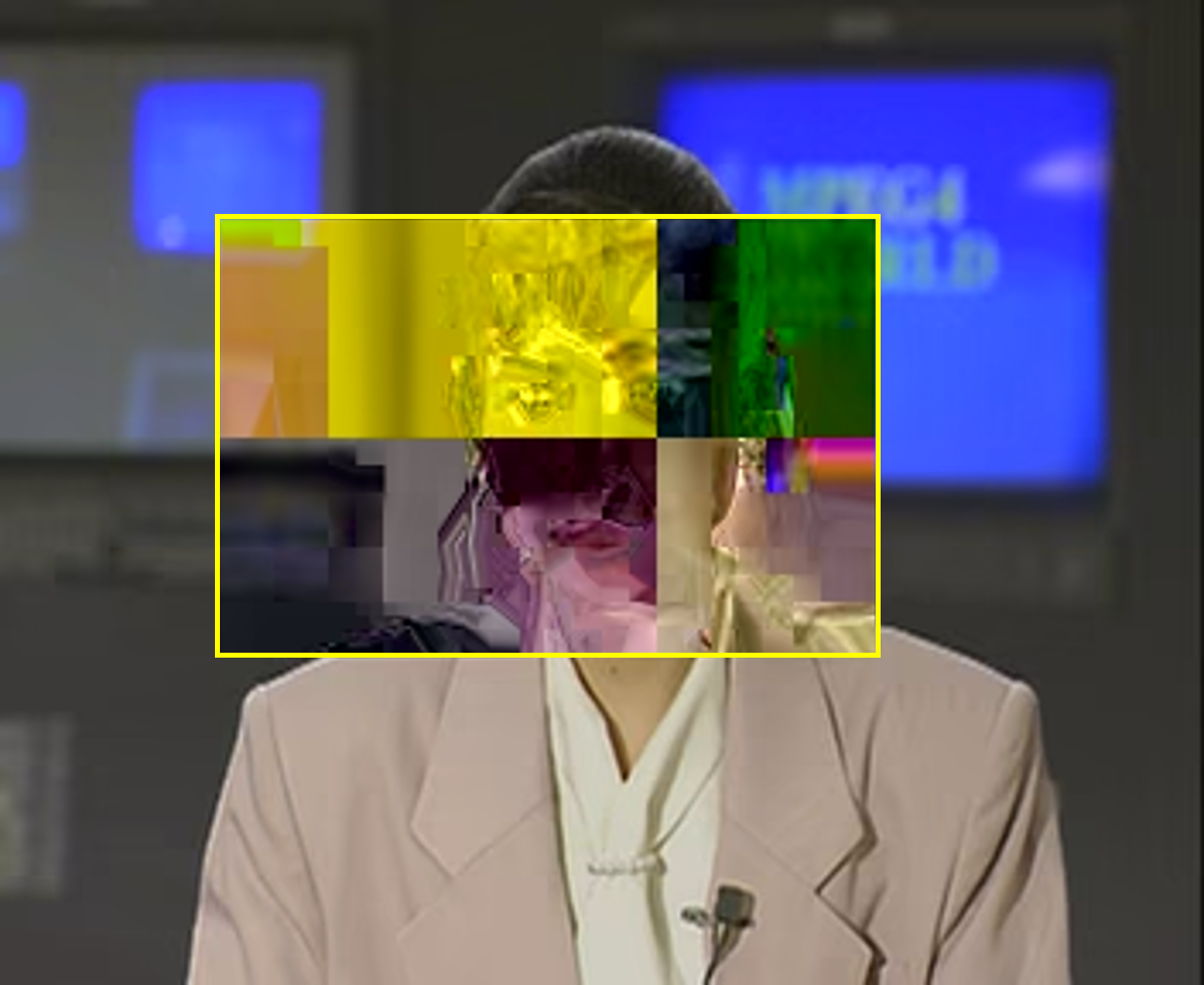}}
\hfill
\subfloat[Zhang]{
\includegraphics[width=0.150\textwidth]{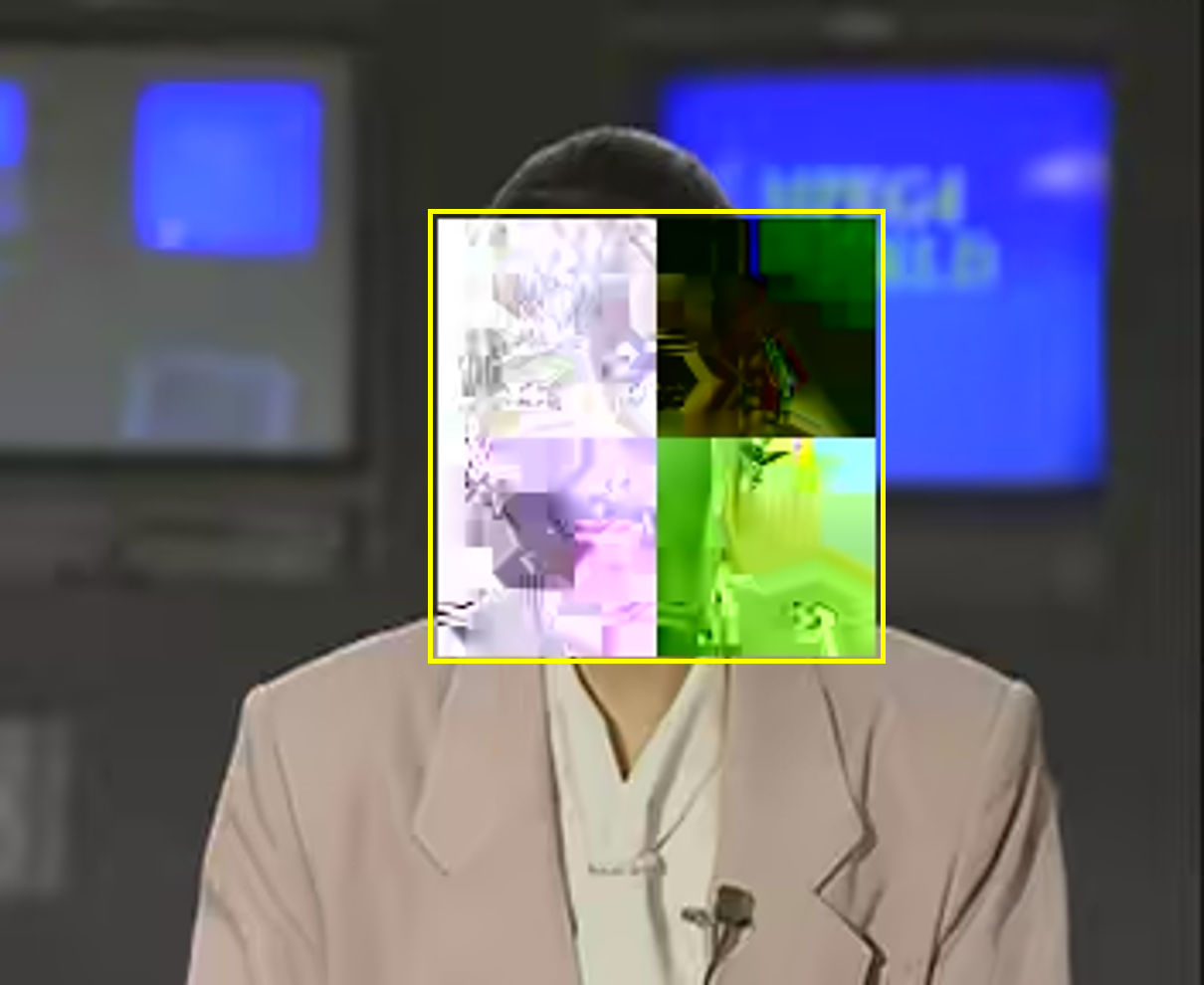}}
\hfill
\subfloat[Yu]{
\includegraphics[width=0.150\textwidth]{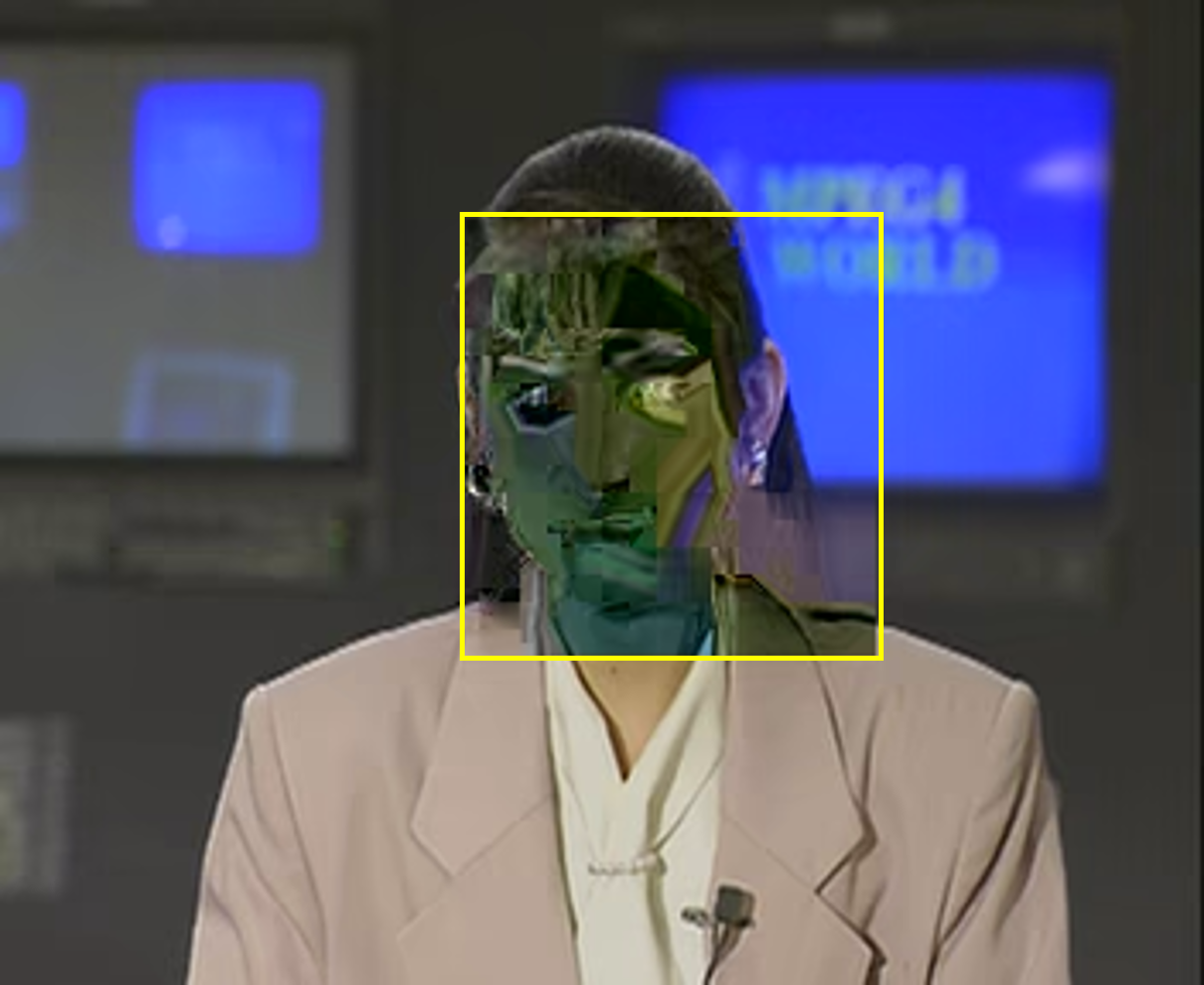}}
\hfill
\subfloat[Amael]{
\includegraphics[width=0.150\textwidth]{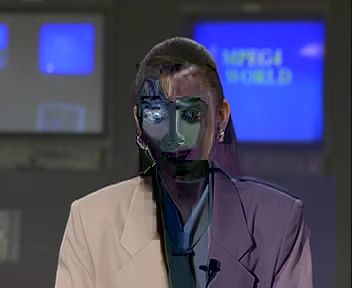}}
\hfill
\subfloat[Ours]{
\includegraphics[width=0.150\textwidth]{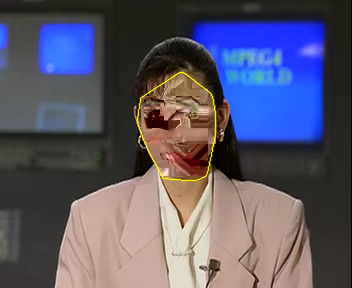}}\\[2pt]
\hfill

\subfloat[Original]{
\includegraphics[width=0.150\textwidth]{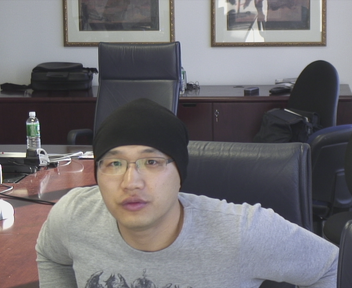}}
\hfill
\subfloat[Taha]{
\includegraphics[width=0.150\textwidth]{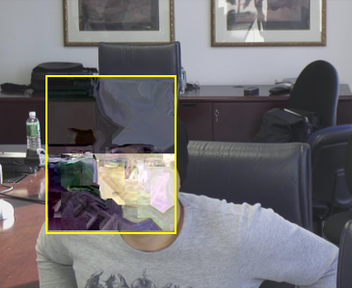}}
\hfill
\subfloat[Zhang]{
\includegraphics[width=0.150\textwidth]{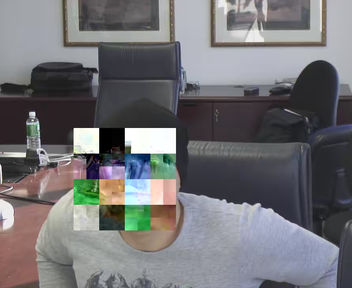}}
\hfill
\subfloat[Yu]{
\includegraphics[width=0.150\textwidth]{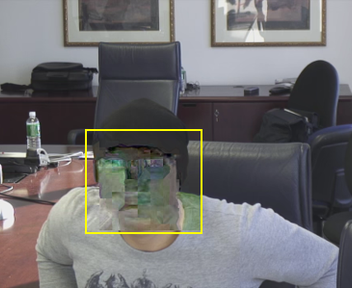}}
\hfill
\subfloat[Amael]{
\includegraphics[width=0.150\textwidth]{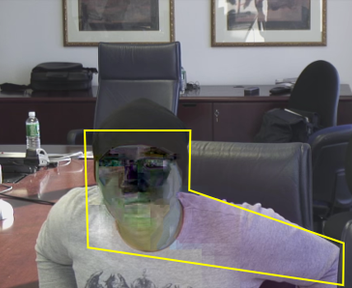}}
\hfill
\subfloat[Ours]{
\includegraphics[width=0.150\textwidth]{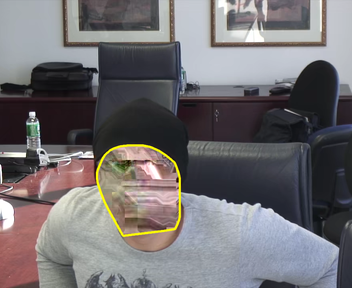}}\\[2pt]
\hfill

\caption{Visual comparison of ROI encryption results on Akiyo and vidyo4 sequence. Each row corresponds to one sequence: the original frame, the results encrypted by ~\cite{taha2018end},~\cite{zhang2025visual},~\cite{yu2023coding},~\cite{11323309}, and the proposed scheme.}
\label{fig:detail}
\end{figure*}

The comparison of ROI perturbation effects includes two main parts: subjective visual comparison and objective index comparison. The following is a detailed analysis of this part.
\paragraph{Comparison of Subjective Vision.}
Figure~\ref{fig:detail} shows a comparison of the perturbation effects on Akiyo and vidyo4 sequence using five different ROI encryption schemes at QP=24. More sequence comparison figures can be found in the section 6 of \textbf{supplementary materials}. As shown in the figure, tile-level ROI encryption schemes \cite{taha2018end,zhang2025visual} are limited by the fixed and coarse rectangular partitions used in tile coding. Because the encryption granularity cannot adapt to the ROI shape, these schemes introduce substantial encryption redundancy. \cite{yu2023coding} mitigate encryption diffusion through CTU-based isolation encryption with a fixed size of 64$\times$64, but the resulting visual effect remains similar to that of tile-level ROI encryption. In contrast, \cite{11323309}’s lightweight encryption strategy does not explicitly address encryption diffusion. Although it targets the smallest syntactic elements related to diffusion, noticeable diffusion effects still occur in most cases. By contrast, the proposed CU-level encryption scheme can follow the natural contours of the ROI at a finer granularity while effectively preserving background whiteboard content by introducing prompt segmentation and $8\times8$ CU-level mapping. This design protects ROI privacy while retaining critical contextual information for viewers, making it suitable for scenarios that require a balance between privacy protection and behavioral analysis.

\paragraph{Comparison of Objective Indicators.}
Table~\ref{obj_indicators} shows the average values of various indicators for the 35 video sequences under five encryption schemes.
\begin{table}[t]
\centering
\small
\renewcommand{\arraystretch}{1.2}
\setlength{\tabcolsep}{2.5pt}

\begin{tabular}{c|cc|ccc}
\toprule
\multirow{2}{*}{Metrics} &
\multicolumn{2}{c|}{Tile-ROI Encryption} &
\multicolumn{3}{c}{CU-ROI Encryption} \\
\cmidrule(lr){2-3}\cmidrule(lr){4-6}

&
Taha &
Zhang &
Yu &
Amael &
Ours \\

\midrule

PSNR
& 11.54 & 10.36 & 12.21 & 13.94 & \cellcolor{myblue}13.29 \\

SSIM
& 0.414 & 0.338 & 0.421 & 0.531 & \cellcolor{myblue}0.485 \\

EDR
& 0.881 & 0.906 & 0.892 & 0.703 & \cellcolor{myblue}0.827 \\

\makecell{Information\\entropy}
& 7.02 & 7.41 & 7.08 & 6.56 & \cellcolor{myblue}7.11 \\

NPCR
& 99.45\% & 99.57\% & 99.55\% & 98.82\% & \cellcolor{myblue}98.93\% \\

UACI
& 28.62\% & 32.92\% & 27.51\% & 26.04\% & \cellcolor{myblue}26.06\% \\

Bitrate Change
& 2.91\% & 8.15\% & 2.11\% & 1.02\% & \cellcolor{myblue}9.31\% \\

\bottomrule
\end{tabular}
\caption{Comparison of objective metrics of ROI for different encryption strategies}
\label{obj_indicators}
\end{table}
The experimental results show that the objective indicator results of the proposed CU-level ROI encryption schemes are weaker than those of existing ROI encryption schemes. This phenomenon is caused by the inherent mechanism of fine-grained encryption, and a detailed theoretical analysis is provided in the section 3 of \textbf{supplementary materials}. However, both the objective metrics and the subjective visual results demonstrate that the proposed scheme introduces sufficient encryption distortion to effectively protect the security of ROI.

\paragraph{Subjective Evaluation Questionnaire Survey.}
To evaluate the subjective perceptual effect of the proposed scheme, we conducted a five-level Likert-scale questionnaire as shown in Table~\ref{evaluation} and invited 120 participants from diverse professional backgrounds to subjectively evaluate the visual differences between each group of frames. The questionnaire focused on two aspects that are directly related to ROI privacy protection: ROI disturbance and non-ROI background preservation. For question 1, participants rated whether the sensitive ROI remained identifiable after encryption, where a higher score indicated stronger privacy protection. For question 2, participants rated whether the background region outside the ROI remained visually clear and usable, where a higher score indicated better preservation of contextual information. The subjective scores were averaged across all participants and test sequences to evaluate whether the proposed method could conceal sensitive ROI content while maintaining the visibility of non-sensitive regions. Table~\ref{score} reports the mean scores for both questionnaire items.

\begin{table}[h]
\centering
\small
\centering 
\setlength{\tabcolsep}{6pt}
\begin{tabular}{ccc} 
\toprule
\multirow{2}{*}{Score} & \multicolumn{2}{c}{Description} \\ 
\cmidrule{2-3} 
 & Question 1 & Question 2\\ 
\midrule
1 & Clearly identifiable  & Severely degraded   \\
2 & Mostly identifiable & Noticeably degraded   \\
3 & Partially obscured & Moderately preserved  \\
4 & Difficult to identify & Well preserved  \\
5 & Completely unrecognizable & Clearly preserved   \\ 
\bottomrule
\end{tabular}
\caption{Subjective rating scale for privacy-utility evaluation}
\label{evaluation}
\end{table}

\begin{table}[h]
\centering
\small
\centering 
\setlength{\tabcolsep}{15pt}
\begin{tabular}{ccc} 
\toprule
\multirow{2}{*}{Group} & \multicolumn{2}{c}{Average Score} \\ 
\cmidrule{2-3} 
 & Question 1 & Question 2\\ 
\midrule
Group A & 4.725 & 4.458   \\
Group B & 4.267 & 4.306   \\
Group C & 4.272 & 4.450  \\
Group D & 4.795 & 4.681  \\ 
Group E & 4.637 & 4.500  \\
Group F & 4.717 & 4.820  \\
Group G & 4.847 & 4.872  \\
\bottomrule
\end{tabular}
\caption{Average scores for different sequences}
\label{score}
\end{table}
As shown in the table, Most group-level mean scores exceeded 4.3, indicating that participants generally found the encrypted ROIs difficult to identify and the non-ROI backgrounds visually clear.

\subsection{Cryptographic Security Analysis}
We further evaluated the proposed scheme through formal key-space analysis, one-bit key-sensitivity testing, known- and chosen-plaintext experiments, NIST SP 800-22 randomness testing, and tamper-detection experiments. All results satisfied the predefined criteria, supporting the confidentiality, integrity, and cross-session security of the scheme under the stated threat model. Detailed settings, evaluation metrics, experiment results, and reproduction procedures are provided in section 4 of \textbf{supplementary materials}.
\begin{table*}[t!]
\centering
\small
\setlength{\tabcolsep}{6pt}
\begin{tabular}{lccccc} 
\toprule
\multirow{2}{*}{Objective Indicators} & \multicolumn{5}{c}{Non-ROI} \\ 
\cmidrule{2-6} 
 & Original & Without isolation & With PCM & With MV restriction  & With PCM+restriction\\ 
\midrule
PSNR & 42.92 & 32.86 &  40.18 &  35.29 &  42.92  \\
SSIM & 0.998 & 0.982 &  0.992 &  0.988 &  0.998  \\
EDR & 0.212 & 0.237 &  0.217 &  0.226 &  0.212  \\
Information entropy & 6.18 & 6.19 &  6.19 &  6.19 &  6.18  \\
NPCR & 73.34\% & 76.52\% &  74.46\% &  75.81\% &  73.34\%  \\ 
UACI & 0.554\% & 1.140\% &  0.971\% &  0.692\% &  0.554\% \\ 
Bitrate change & 0 & 2.02\% &  6.62\% &  2.28\% &  9.31\% \\
\bottomrule
\end{tabular}
\caption{Comparison of objective metrics with and without isolation}
\label{tab:Non-ROI}
\end{table*}
\subsection{Ablation Study}
To quantitative evaluate the effectiveness of the proposed diffusion isolation mechanism, we conducted ablation experiments and present the results of objective metrics for non-ROI regions with and without diffusion isolation (PCM and MV restriction). The testing results are shown in Table~\ref{tab:Non-ROI}. The original column in the table indicates the quality of the non-ROI region in unencrypted frames, whereas the remaining columns correspond to encryption without isolation, only PCM-based isolation, only MV restriction, and their combination, respectively. Since all quality metrics are calculated only over non-ROI pixels, the results closer to the original column indicates more effective suppression of encryption distortion propagation.

These results show that the distortion introduced in encrypted ROI CUs propagates into the unencrypted region when no isolation mechanism is applied. PCM-based isolation considerably mitigates this spatial distortion propagation. Nevertheless, the non-ROI quality does not fully return to the reference level, indicating that PCM-based isolation alone cannot eliminate propagation caused by inter frame prediction. MV restriction can suppress temporal propagation, but it cannot fully address the spatial propagation associated with intra prediction and reconstruction dependencies. Therefore, PCM-based isolation and MV restriction are complementary, the combined PCM and MV restriction scheme achieves the complete non-ROI preservation. It must be acknowledged that using the PCM mode inevitably leads to an increase in bit rate. In addition, we have further provided visualization figures of the ablation experiments in section 5 of \textbf{supplementary materials}.

\section{Conclusions}\label{sec6}
This paper proposes a prompt segmentation-guided CU-level ROI encryption algorithm. In response to the problems of over encryption  in existing CU-level ROI encryption schemes, a more adaptive CU-level ROI encryption strategy is designed. In response to encryption diffusion, we propose a dual diffusion isolation strategy based on PCM mode and MV restriction, effectively blocking pixel dependent propagation between ROI and non-ROI, thereby improving the independence and controllability of region encryption. The experimental results show that the algorithm can achieve more detailed regional perturbation effects while ensuring the safety of ROI. Future work will attempt to introduce a combination of chaos models and coefficient perturbation mechanisms to enhance the randomness and unpredictability, and improve the robustness of regional perturbations.

\bibliography{aaai2027}

\end{document}